\documentclass{amsart} 
\RequirePackage{fix-cm}
\RequirePackage{amsmath}

\usepackage[english]{babel}
\usepackage[utf8x]{inputenc}
\usepackage{amsmath}
\usepackage{graphicx}
\usepackage[colorinlistoftodos]{todonotes}
\usepackage{amsfonts}
\usepackage{color}
\usepackage{soul}
\usepackage{listings}
\usepackage{epstopdf}
\usepackage{pdfpages}
\usepackage[a4paper, total={6in, 8in}]{geometry}
\usepackage{hyperref}
\usepackage[section]{placeins}

\usepackage{color}

\DeclareMathOperator{\median}{median}
\DeclareMathOperator{\corr}{corr}

\newtheorem{OpenProb}{Open Problem}

\begin{document}

\title{Revised dynamics of the Belousov-Zhabotinsky reaction model}

\author[J. Nagyov\'{a}]{Judita Nagyov\'{a}}
\address{IT4Innovations, V\v{S}B - Technical University of Ostrava,
17. listopadu 15/2172, 708 33 Ostrava, Czech Republic\\
Department of Applied Mathematics, V\v{S}B - Technical University of Ostrava,
17. listopadu 15/2172, 708 33 Ostrava, Czech Republic}
\email{judita.nagyova.st@vsb.cz}
					
\author[B. Jans\'{i}k]{Branislav Jans\'{i}k} 
\address{IT4Innovations, V\v{S}B - Technical University of Ostrava,
17. listopadu 15/2172, 708 33 Ostrava, Czech Republic}
\email{branislav.jansik@vsb.cz}
       
\author[M. Lampart]{Marek Lampart}
\address{IT4Innovations, V\v{S}B - Technical University of Ostrava,
17. listopadu 15/2172, 708 33 Ostrava, Czech Republic\\
Department of Applied Mathematics, V\v{S}B - Technical University of Ostrava,
17. listopadu 15/2172, 708 33 Ostrava, Czech Republic}
\email{marek.lampart@vsb.cz}

\begin{abstract}
The main aim of this paper is to detect dynamical properties of the Györgyi-Field model of the Belousov-Zhabotinsky chemical reaction.
The corresponding three-variable model given as a set of nonlinear ordinary differential equations depends on one parameter, the flow rate. 
As certain values of this parameter can give rise to chaos, the analysis was performed in order to identify different dynamics regimes.
Dynamical properties were qualified and quantified using classical and also new techniques. 
Namely, phase portraits, bifurcation diagrams, the Fourier spectra analysis, the 0-1 test for chaos, and approximate entropy. 
The correlation between approximate entropy and the 0-1 test for chaos was observed and described in detail. 
Moreover, the three-stage system of nested subintervals of flow rates, for which in every level the 0-1 test for chaos and approximate entropy was computed, is showing the same pattern. 
The study leads to an open problem whether the set of flow rate parameters has Cantor like structure.

\end{abstract}

\maketitle

\section{Introduction}

The Belousov-Zhabotinsky chemical reaction (BZ reaction), originally discovered in the 1950s by Boris P. Belousov \cite{origin}, 
is an example of oscillating chemical reaction which can be maintained far from equilibrium by an internal source of energy \cite{oreg} resulting in a nonlinear chemical oscillator exhibiting different dynamical regimes. 
Later on, the chemical mechanism of the reaction was described in \cite{Field1972}, what is commonly called the FKN mechanism. 

There are many mathematical models representing different aspects of the BZ reaction. 
For example, the Brusselator, Oregonator and Györgyi-Field are three mathematical models for a type of autocatalytic reaction -- like the BZ reaction. 

The Oregonator model is the result of a quantitative kinetic analysis of oscillations in the BZ reaction by Field and Noyes in 1974 \cite{doi:10.1063/1.1681288} and it is a simplified version of the mechanism developed by Field, Körös and Noyes (FKN mechanism) \cite{origin}. 

The Brusselator model, a theoretical model for a type of autocatalytic reaction, was proposed by I. Prigogine and his collaborators \cite{glansdorff1971thermodynamic}.

Finally, the Györgyi-Field model (GF model), describes the reaction taking place in a continuous-flow stirred-tank reactor (CSTR) \cite{bz3}, 
by a relatively simple mathematical model (see also \cite{Gyorgyi1991} and \cite{Gyorgyi19912}). This model, for a specific choice of parameters, exhibits chaos (see e.g. \cite{epstein1998introduction} and references therein or main results of this paper), 
on contrary to the Oregonator which has no chaotic solutions \cite{doi:10.1063/1.440418} describing the oscillatory behaviour and pattern formation in the BZ reaction.
The GF model will be taken into consideration for further research in this paper.

\medskip 

In recent decades,  the BZ reaction has been extensively studied by physical chemists on its kinetic behaviour \cite{epstein1998introduction,xx18,xx26} and by mathematicians on the dynamics and patterns of the solutions of the associated mathematical model \cite{xx7,xx16,xx19,doi:10.1063/1.440418}.

More specifically, the research from the theory of dynamical system point of view was done.  
The transition from steady state to quasi-periodic and bursting oscillations, and further on to regular relaxation oscillation via a complicated sequence of alternating periodic and chaotic regimes were done by simulations in \cite{NOSKOV199482}. 
The results of computer experiments on information processing in a hexagonal array of vesicles filled with BZ solution in a sub-excitable mode were introduced by \cite{ADAMATZKY2011779}.
The discretized version of BZ reaction models was also researched. 
E.g. in \cite{Kang2005},  the dynamics of the local map is discussed, the set of trajectories that escape to infinity as well as analyze the set of bounded trajectories – the Julia set of the system.
The evidence of chaos was also demonstrated in an experimental way by dozens works e.g. \cite{doi:10.1063/1.435267,Yamazaki,doi:10.1063/1.438487}.

Despite a large number of results in the given area, it is possible to apply new methods to the given BZ reaction model and to obtain new very interesting results that better characterize the trajectory behaviour depending on the choice of state parameters showing properties of the parameter space.

\medskip

This work focuses on the characterization of dynamical properties of the GF model \cite{bz3} depending on the flow rate, denoted by $k_f$. 
The qualitative and quantitative characterization of the dynamics regimes is mainly done using the 0-1 test for chaos and approximate entropy. 
Recall that these tools were applied in \cite{xx} to the two-dimensional coupled map lattice model of the Lagrangian type, which is a discrete version of the BZ reaction.
These tools were applied to the huge simulation data that was performed on the Salomon supercomputer 
at IT4Innovations National Supercomputing Center located in Ostrava, Czech Republic.

\medskip

The paper is organized as follows: in Section \ref{sec:model} the model is introduced, followed by its mathematical model in Section \ref{sec:mathmodel}. 
The main results obtained by phase portraits, the Fourier spectra analysis, the approximate entropy, and the 0-1 for chaos, are contained in Section \ref{sec:results}. 
Finally, the outcomes are summarized in Section \ref{sec:concl}.

\section{The Györgyi-Field reaction model}\label{sec:model}

The GF model of the BZ reaction develops a description of the reaction in terms of a set of differential equations containing only three variables. 
In common with experiments, the GF model shows both regular, intermittent and chaotic behavior. 
While remaining close to a real chemical system, it is sufficiently simple to allow detailed mathematical analysis \cite{bz3}. The mechanism of the reaction is defined by the set of the following equations (\ref{modelchem}):

\begin{subequations}\label{modelchem}
\begin{align}
Y+X+H &\rightarrow 2V \label{chem1}\\
Y+A+2H &\rightarrow V+H \label{chem2}\\
2X &\rightarrow V \label{chem3}\\
0.5 X+A+H &\rightarrow X+Z \label{chem4}\\
X+Z &\rightarrow 0.5 X \label{chem5}\\
V+Z &\rightarrow Y \label{chem6}\\
Z+M &\rightarrow \label{chem7} 
\end{align}
\end{subequations}
where the corresponding chemical components are: \(Y=\) Br\textsuperscript{-}, \(X=\) HBrO\textsubscript{2}, \(Z=\) Ce$^{4+}$, \(V=\) BrCH(COOH)\textsubscript{2} or BrMA, \(A=\) BrO\textsubscript{3}\textsuperscript{-}, \(H=\) H\textsuperscript{+}, \(M=\) CH\textsubscript{2}(COOH)\textsubscript{2}. The concentrations of the main reactants \(A\), \(H\), \(M\), and the total concentration of cerium ions \(C\) are summarized in Table \ref{tab1}.

\begin{table}
\caption{Rates and rate constants of the GF model chemical scheme.}
\label{tab3}
\begin{tabular}{c l l}
  \hline\noalign{\smallskip}
	Reaction equation & Rate \(r_i\) & Rate constant \(k_i\)\\
	\noalign{\smallskip}\hline\noalign{\smallskip}
	\eqref{chem1} & \(r_1 = k_1 HYX\) & \(k_1 = 4.0 \times 10^6 \,M^{-2}s^{-1}\)\\
	\hline\noalign{\smallskip}
	\eqref{chem2} & \(r_2 = k_2 AH^2 Y\) & \(k_2 = 2.0 \,M^{-3}s^{-1}\)\\
	\hline\noalign{\smallskip}
	\eqref{chem3} & \(r_3 = k_3 X^2 \) & \(k_3 = 3000 \,M^{-1}s^{-1}\)\\
	\hline\noalign{\smallskip}
	\eqref{chem4} & \(r_4 = k_4 A^{0.5}H^{1.5}(C-Z)X^{0.5} \) & \(k_4 = 55.2 \,M^{-2.5}s^{-1}\)\\
	\hline\noalign{\smallskip}
	\eqref{chem5} & \(r_5 = k_5 XZ \) & \(k_5 = 7000 \,M^{-1}s^{-1}\)\\
	\hline\noalign{\smallskip}
	\eqref{chem6} & \(r_6 = \alpha k_6 ZV \) & \(k_6 = 0.09 \,M^{-1}s^{-1}\)\\
	\hline\noalign{\smallskip}
	\eqref{chem7} &\(r_7 = \beta k_7 MZ \) & \(k_7 = 0.23 \,M^{-1}s^{-1}\)\\
	\noalign{\smallskip}\hline
\end{tabular}
\end{table}

\section{Mathematical model}\label{sec:mathmodel}

A three-variable mathematical model of the BZ reaction, presented by Györgyi and Field in \cite{bz3}, 
describes the reaction taking place in a CSTR. 
The corresponding set of nonlinear ordinary differential equations contains only three variables, 
while still being able to accurately reproduce the behavior of the BZ reaction observed experimentally \cite{bz3} 
and it is based on a four-variable chemical mechanism \eqref{modelchem}, see \cite{bz3}.

The mathematical model, in its dimensionless form, consists of a set of scaled differential equations:

\begin{subequations}\label{all11}
\begin{align}
\frac{dx}{d\tau}&=T_0 ( -k_1 HY_0 x\tilde{y}+k_2 A H^2 Y_0/X_0 \tilde{y}-2k_3 X_0 x^2+0.5 k_4 A^{0.5} H^{1.5}X_0^{-0.5}(C-Z_0 z)x^{0.5}- \label{11}\\
&\phantom{{}=}-0.5 k_5 Z_0 xz -k_f x)  \notag\\
\frac{dz}{d\tau}&=T_0 \left( k_4 A^{0.5} H^{1.5} X^{0.5} (C/Z_0 -z)x^{0.5}-k_5 X_0 xz -\alpha k_6 V_0 zv-\beta k_7 Mz - k_f z \right) \label{12}\\
\frac{dv}{d\tau}&=T_0 \left( 2k_1 HX_0 Y_0/V_0 x \tilde{y}+k_2 AH^2 Y_0/V_0 \tilde{y}+k_3 X^2_0/V_0 x^2-\alpha k_6 Z_0 zv -k_f v \right) \label{13}
\end{align}
\end{subequations}
where
\(\tau = t/T_0\), \(x = X/X_0\), \(z = Z/Z_0\), \(v = V/V_0\), 
and
\(\tilde{y}= \left(\alpha k_6 Z_0 V_0 zv / \left( k_1 HX_0 x +k_2 AH^2 + k_f \right)\right)/Y_0\)
while \(t\) corresponding to time, \(X\) to HBrO\textsubscript{2}, \(Y\) to Br\textsuperscript{-}, \(Z\) to Ce$^{4+}$, and \(V\) to BrMA.
The rate constants and parameters used in the following computations are given in the Table \ref{tab3} and \ref{tab1}, respectively.

\begin{table}
\caption{Parameters of the investigated system \eqref{all11}.}
\label{tab1}
\begin{tabular}{l}
  \hline\noalign{\smallskip}
	List of parameters\\
	\noalign{\smallskip}\hline\noalign{\smallskip}
	\(A = 0.1\)\\
	\hline
	\(M = 0.25\)\\
	\hline
	\(H = 0.26\)\\
	\hline
	\(C = 0.000833\)\\
	\hline
	\(\alpha = 666.7\)\\
	\hline
	\(\beta = 0.3478\)\\
	\noalign{\smallskip}\hline
\end{tabular}
\end{table}

The behavior of this system depends on the inverse residence time of the CSTR, the flow rate, noted \(k_f\) [$s^{-1}$]. As certain values of this parameter can give rise to chaos, the following analysis was performed in order to identify different dynamics.

\vspace{0.1cm}
\section{Main results}\label{sec:results}

The system of differential equations \ref{all11} was solved numerically in Matlab \cite{Matlab} using the \(ode45\) solver. 
The simulations were done depending on free parameter \(k_f\) ranging from \(3 \times 10^{-4}\) to \(5 \times 10^{-4}\) with \(10^{-7}\) step.
Each simulation was performed for the final time \(\tau = 100\) with a time step \(10^{-4}\). 
To avoid the systems distortions, only the last 20\% of simulations were used for further computations. In all cases, initial conditions were set
$$x_0 =z_0 = v_0 = 1.$$

As a main results, phase diagrams, amplitude frequency spectrums (FFT), and Poincar\'{e} sections were done for relevant choices of the parameter \(k_f\). 
To illustrate changes in dynamical behavior, bifurcation diagrams underlined by the approximate entropy and the 0-1 test for chaos with suitable magnifications to the parameter \(k_f\) were plotted.

Consequently, as a goal of this paper, 
bifurcation diagrams, the approximate entropy, and the 0-1 test for chaos were computed for nested set of parameters \(k_f\).  
The 0-1 test for chaos splits the values of the parameter for which regular (periodic or quasi-periodic) and irregular (chaotic) movements appear, 
while the output of approximate entropy detects increasing complexity of the investigated system \ref{all11}. 

\subsection{Phase diagrams, the Fourier spectra, and bifurcation diagrams}

\begin{figure}
\centering
	\begin{tabular}{l c r}
		\fbox{\includegraphics[scale=0.3]{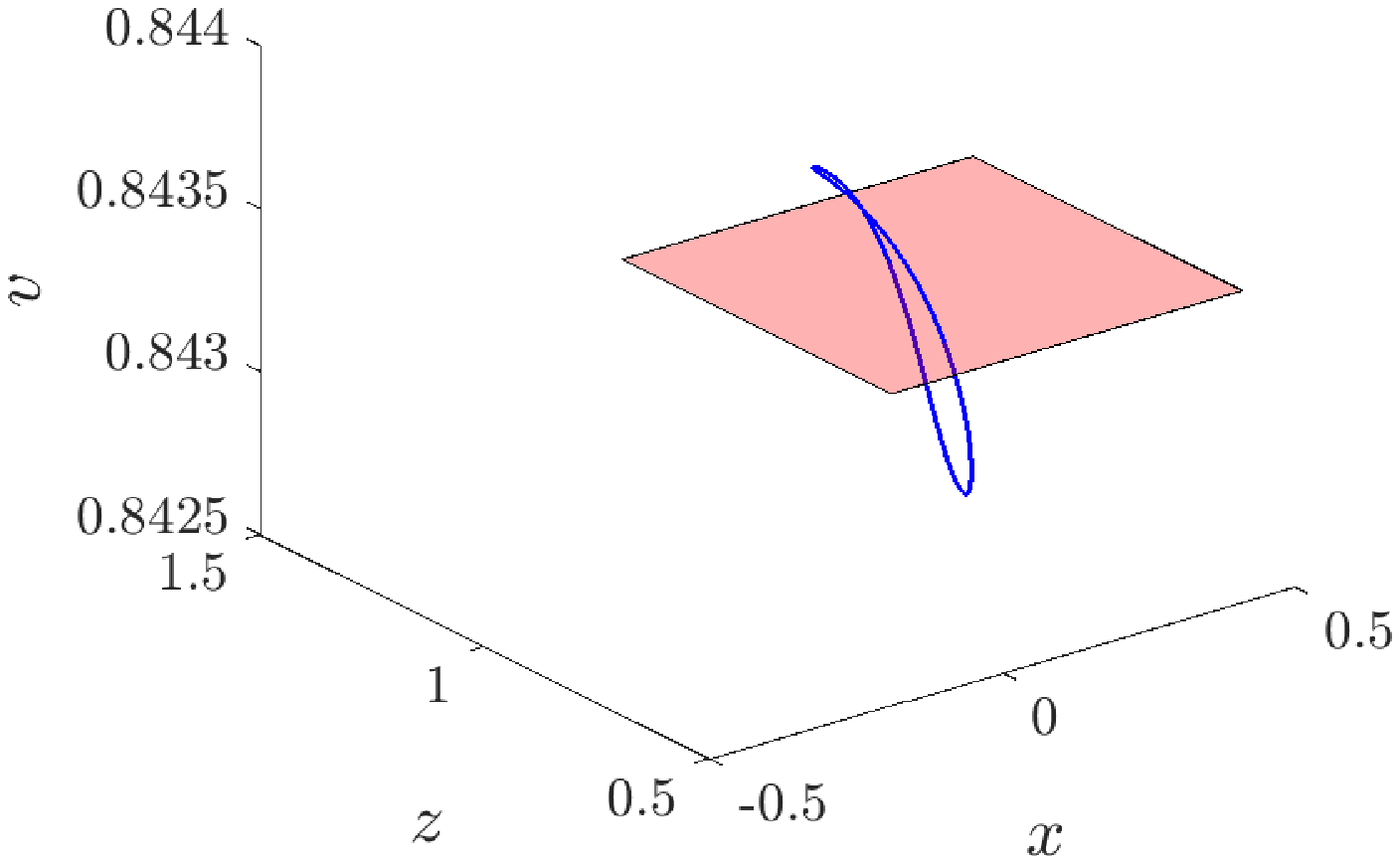}} & \fbox{\includegraphics[scale=0.3]{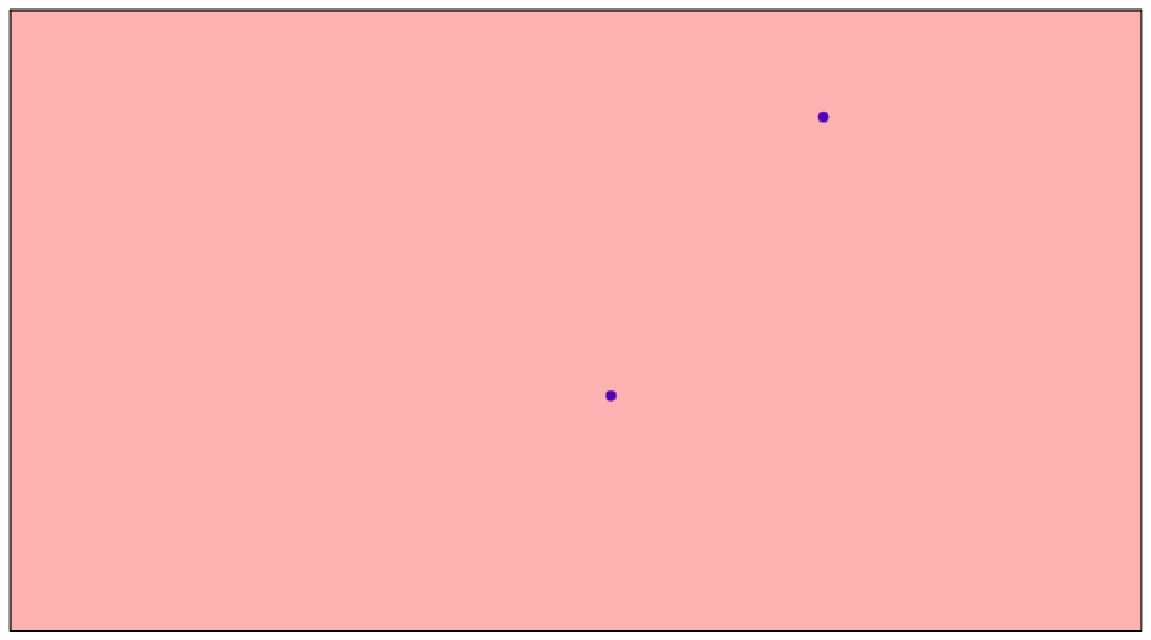}} & \fbox{\includegraphics[scale=0.3]{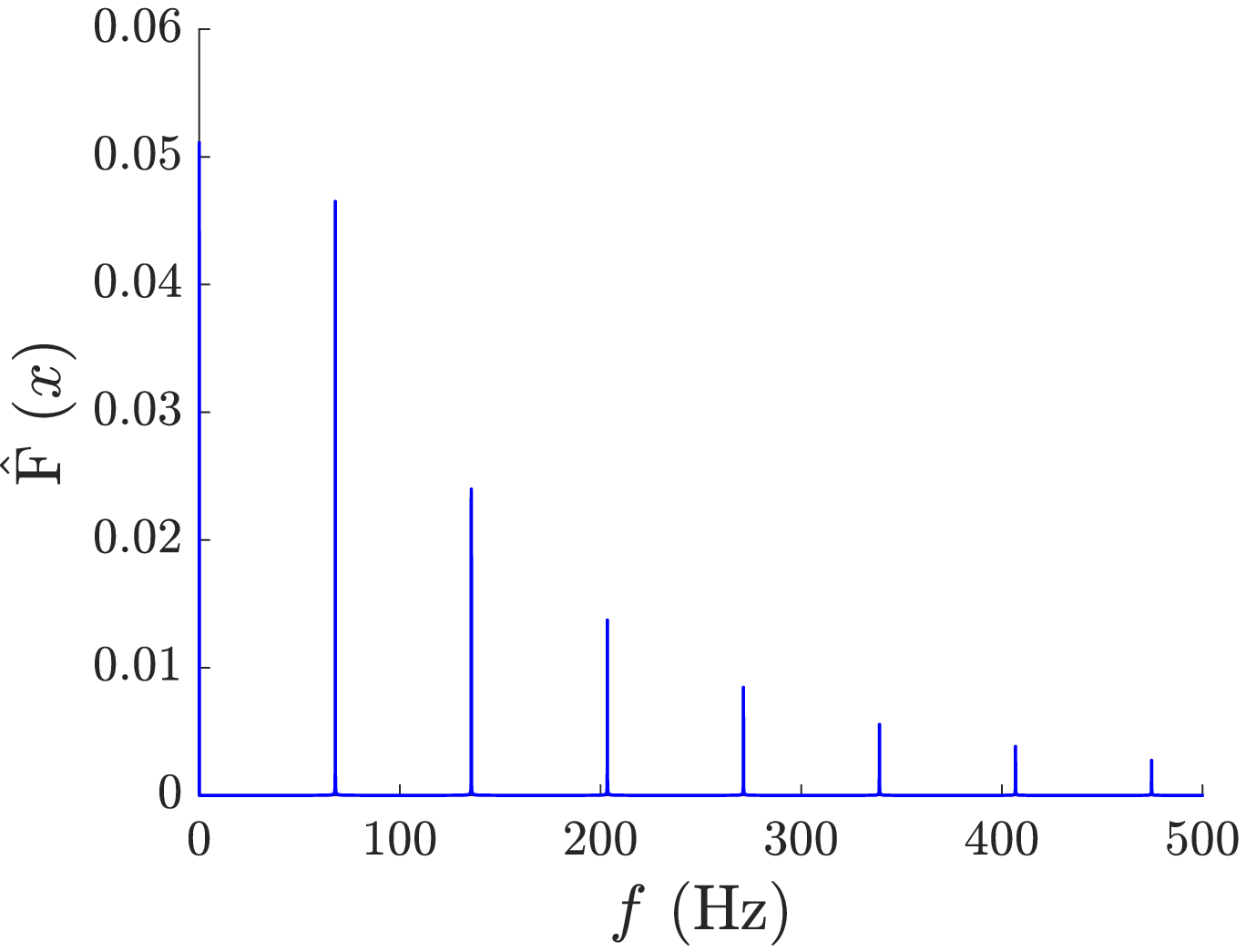}}
	\end{tabular}
	 \caption{Phase diagram with Poincar\'{e} section by plane \(v=0.8434\) and Fourier transform of variable \(x\) for \(k_f=3 \times 10^{-4}\).}
	\label{kf3}
\end{figure}

\begin{figure}
\centering
	\begin{tabular}{l c r}
		\fbox{\includegraphics[scale=0.3]{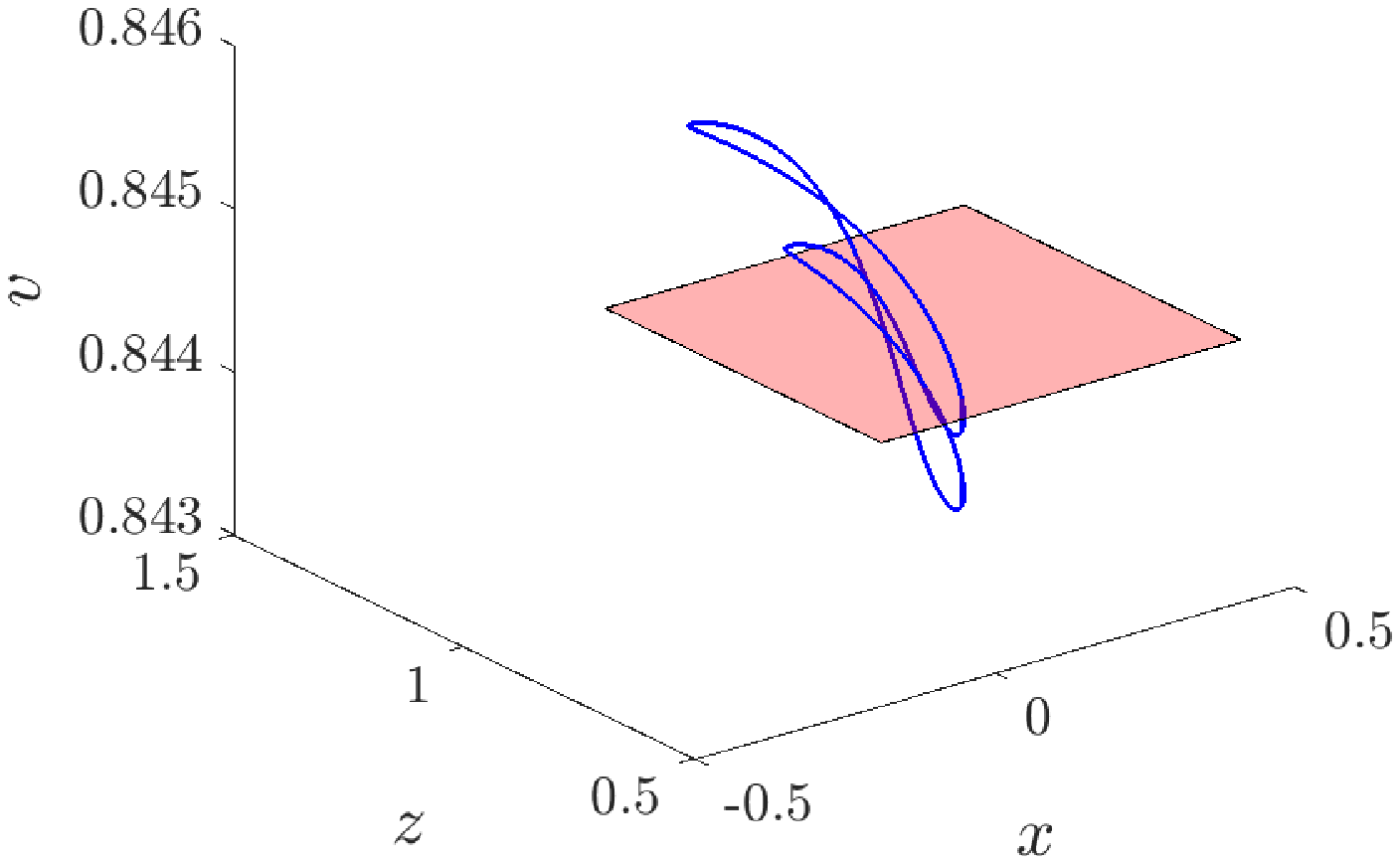}} & \fbox{\includegraphics[scale=0.3]{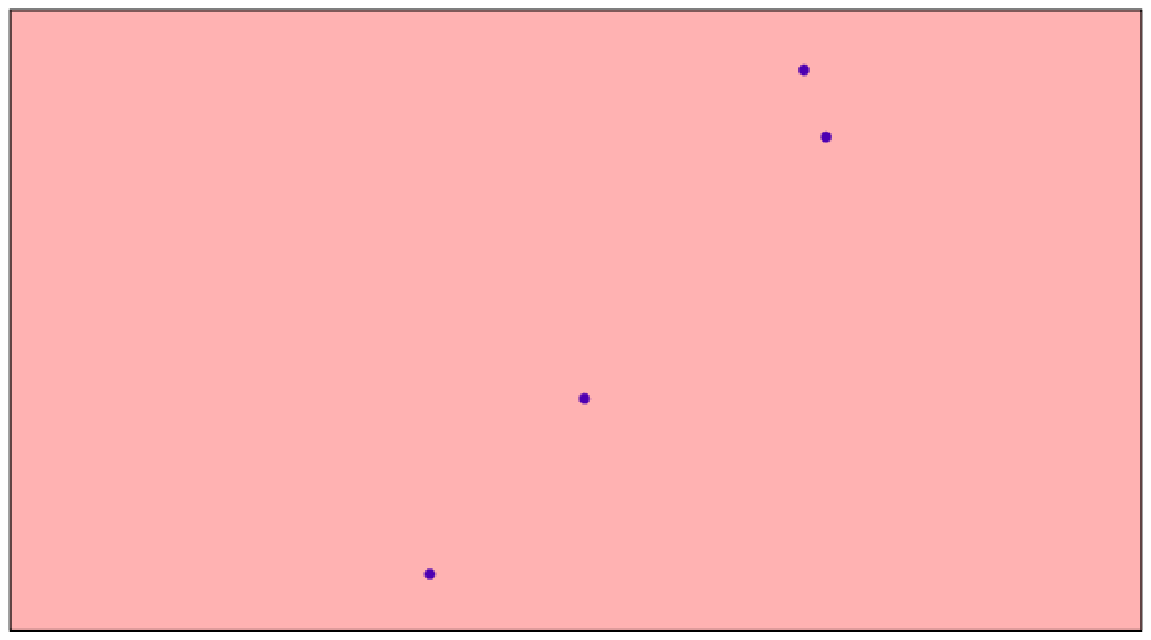}} & \fbox{\includegraphics[scale=0.3]{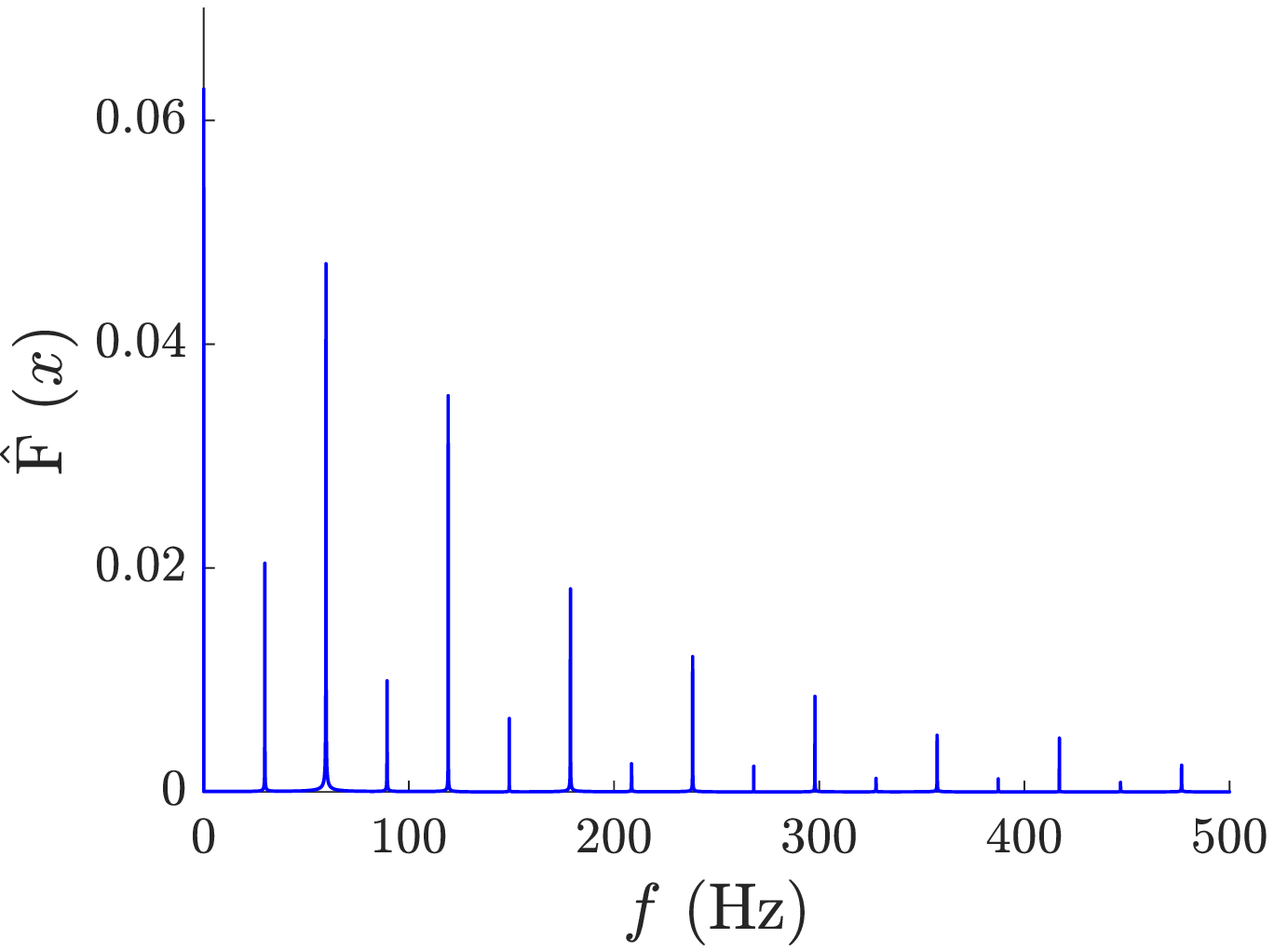}}
	\end{tabular}
	 \caption{Phase diagram with Poincar\'{e} section by plane \(v=0.8445\) and Fourier transform of variable \(x\) for \(k_f=3.2 \times 10^{-4}\).}
	\label{kf32}
\end{figure}

\begin{figure}
\centering
	\begin{tabular}{l c r}
		\fbox{\includegraphics[scale=0.3]{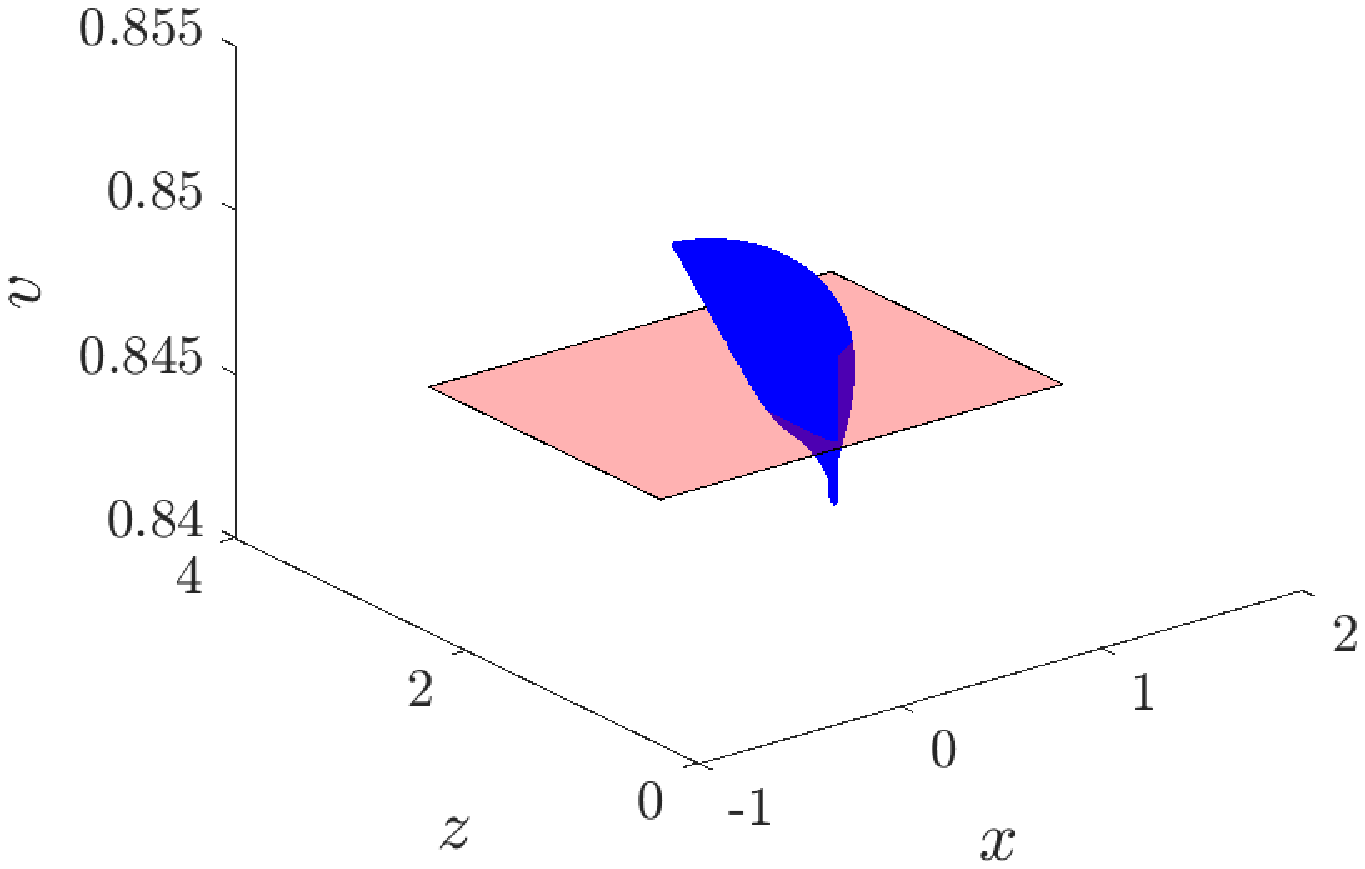}} & \fbox{\includegraphics[scale=0.3]{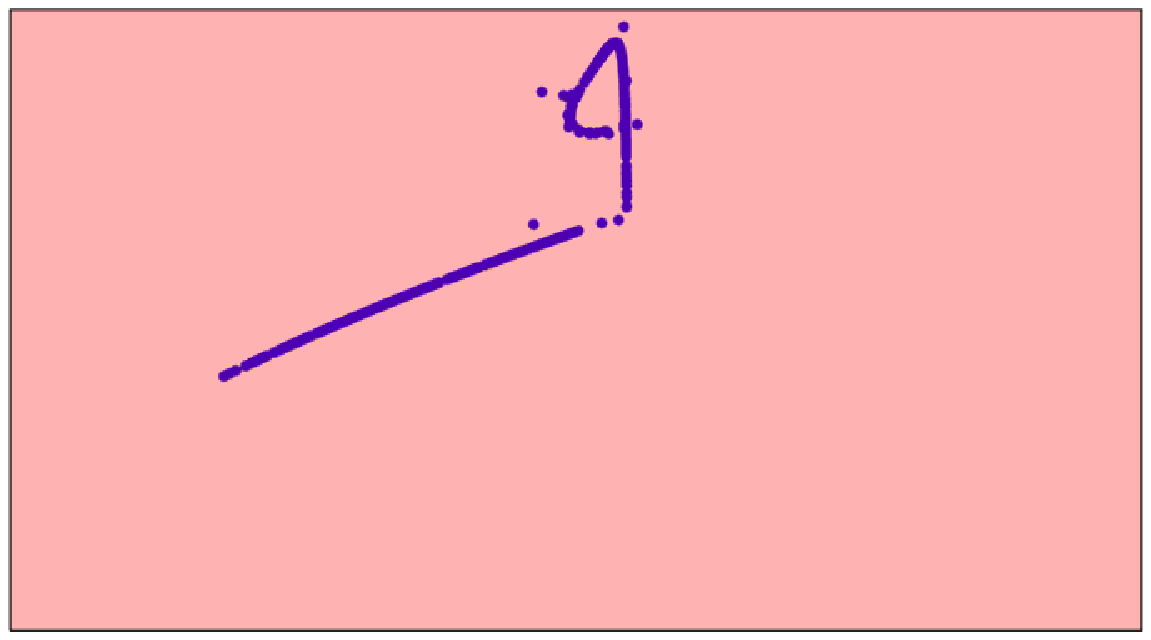}} & \fbox{\includegraphics[scale=0.3]{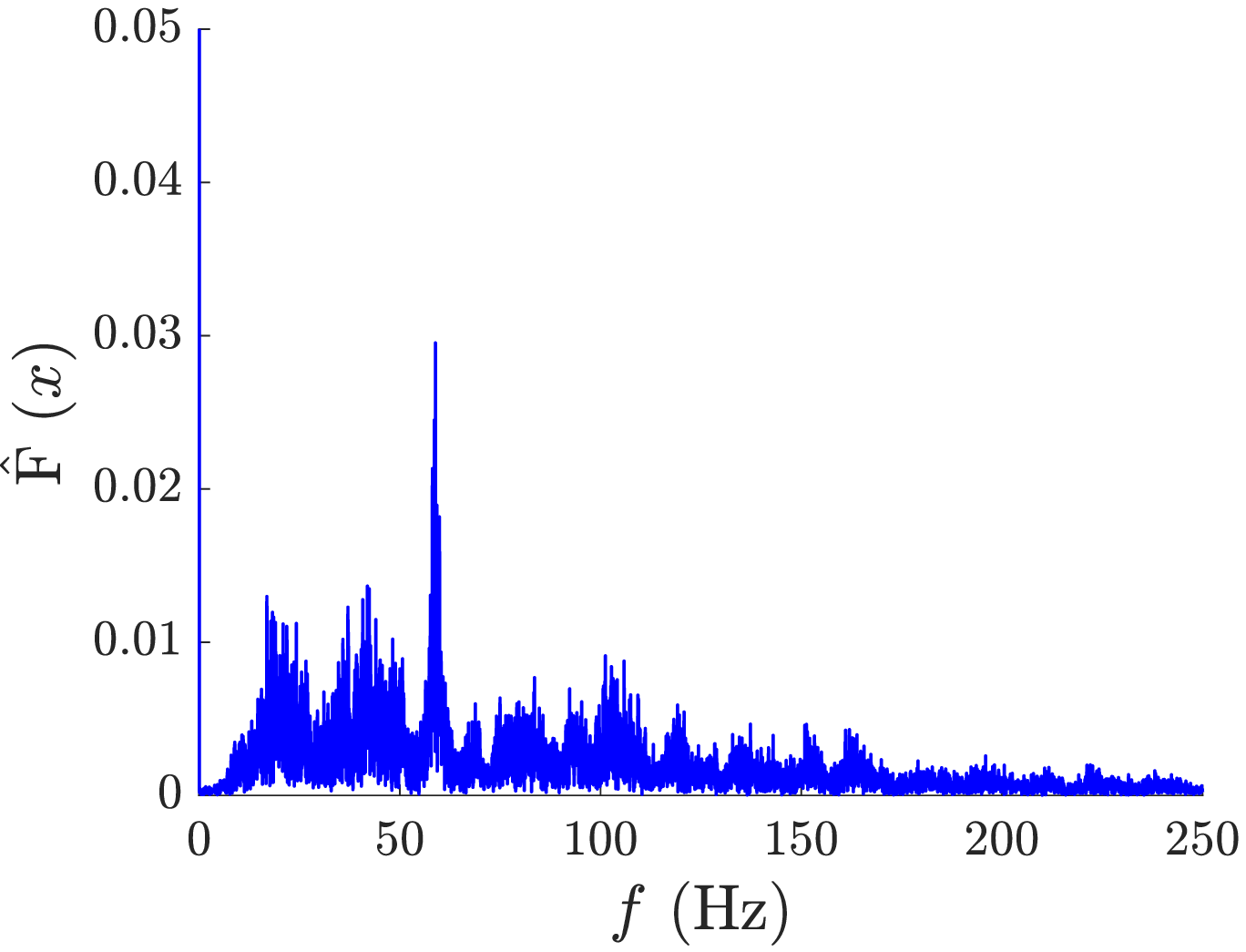}}
	\end{tabular}
	 \caption{Phase diagram with Poincar\'{e} section by plane \(v=0.847\) and Fourier transform of variable \(x\) for \(k_f=3.5 \times 10^{-4}\).}
	\label{kf35}
\end{figure}

Periodic as well as chaotic dynamics was identified in the studied model (\ref{all11}).
For example, in Fig.~\ref{kf3} and Fig.~\ref{kf32}, regular movement is shown by a trivial loop (Fig.~\ref{kf3}) for \(k_f=3\times 10^{-4}\), and a non-trivial loop (Fig.~\ref{kf32}) for \(k_f=3.2\times 10^{-4}\).
Figure \ref{kf35} gives an example of a chaotic trajectory, that is for \(k_f=3.5\times 10^{-4}\).

The Fourier spectra were computed by Fast Fourier transform for \(k_f=3\times 10^{-4}\), \(k_f=3.2\times 10^{-4}\), and \(k_f=3.5\times 10^{-4}\), shown in Figs.~\ref{kf3}, \ref{kf32}, and \ref{kf35} respectively. 
Regular behavior is observable for the first two and chaos in the last case.

In the case of regular movement, in Fig.~\ref{kf3} and Fig.~\ref{kf32}, the Fourier spectra is formed by a number of harmonic frequencies, hence the frequency of the periodic trajectory is computable. Periodic motions of trajectory is also visible in Poincar\'{e} sections.

In the case of chaotic movement, Fig.~\ref{kf35}, the Fourier spectra is formed by a number of harmonic components having the basic, super-harmonic, sub-harmonic, and combination frequencies on which further motions with frequencies forming the sided bands of the dominant frequencies are superposed. Their mutual ratio indicates the irregularity of the motion. The character of this chaotic motion is underlined by the Poincar\'{e} section.

\begin{figure}[h!]
\centering
	\begin{tabular}{l c r}
		\fbox{\includegraphics[scale=0.3]{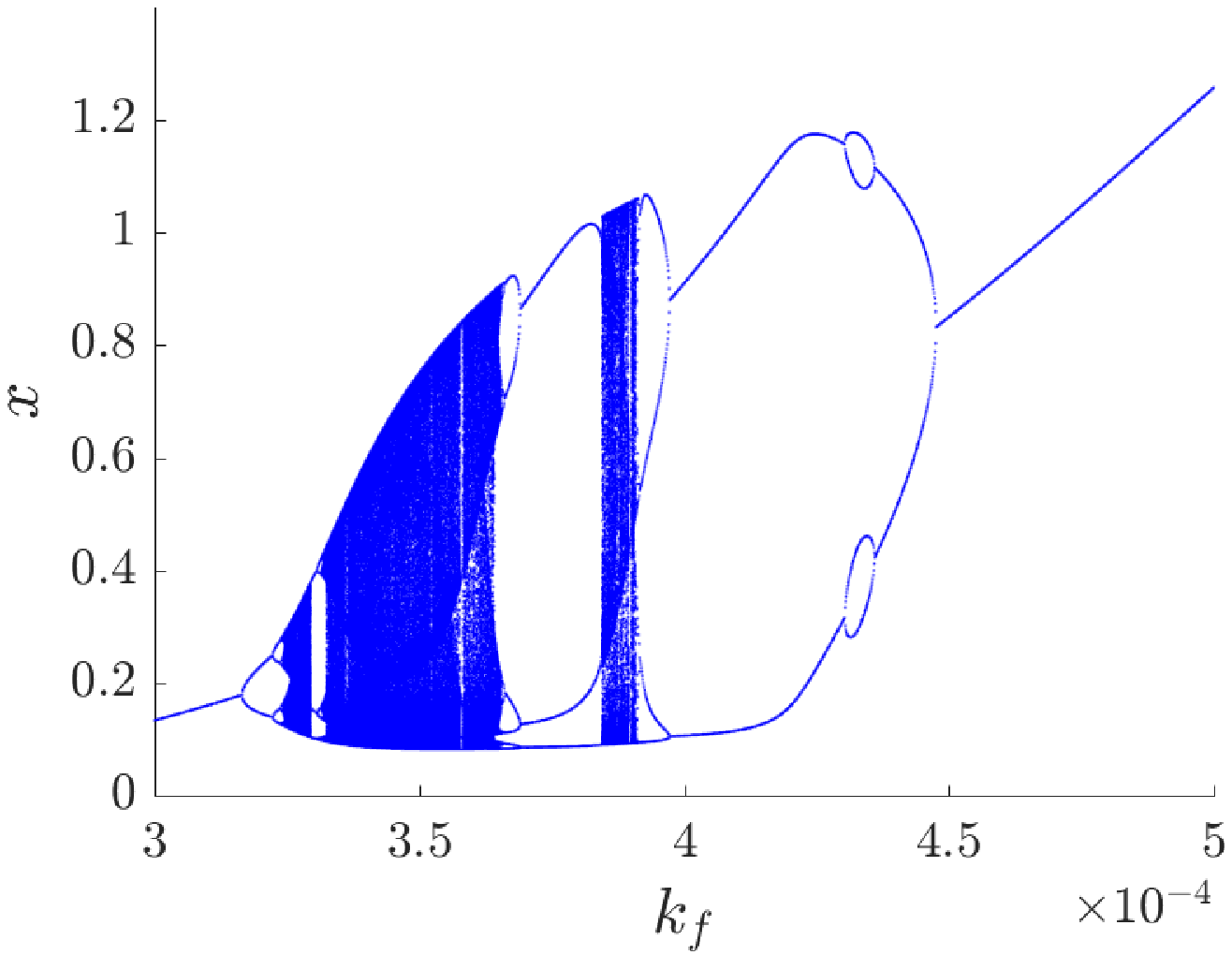}} & \fbox{\includegraphics[scale=0.3]{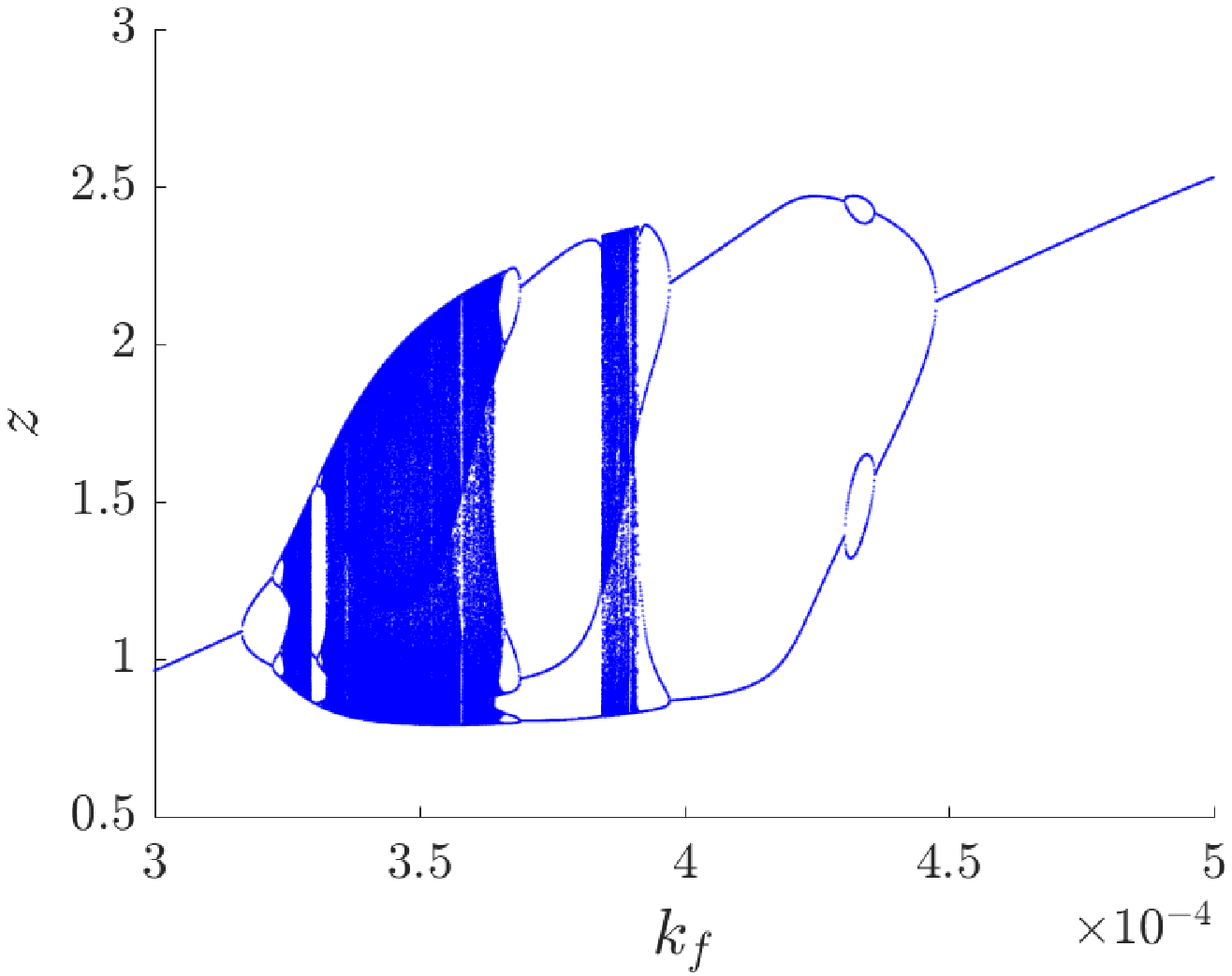}} & \fbox{\includegraphics[scale=0.3]{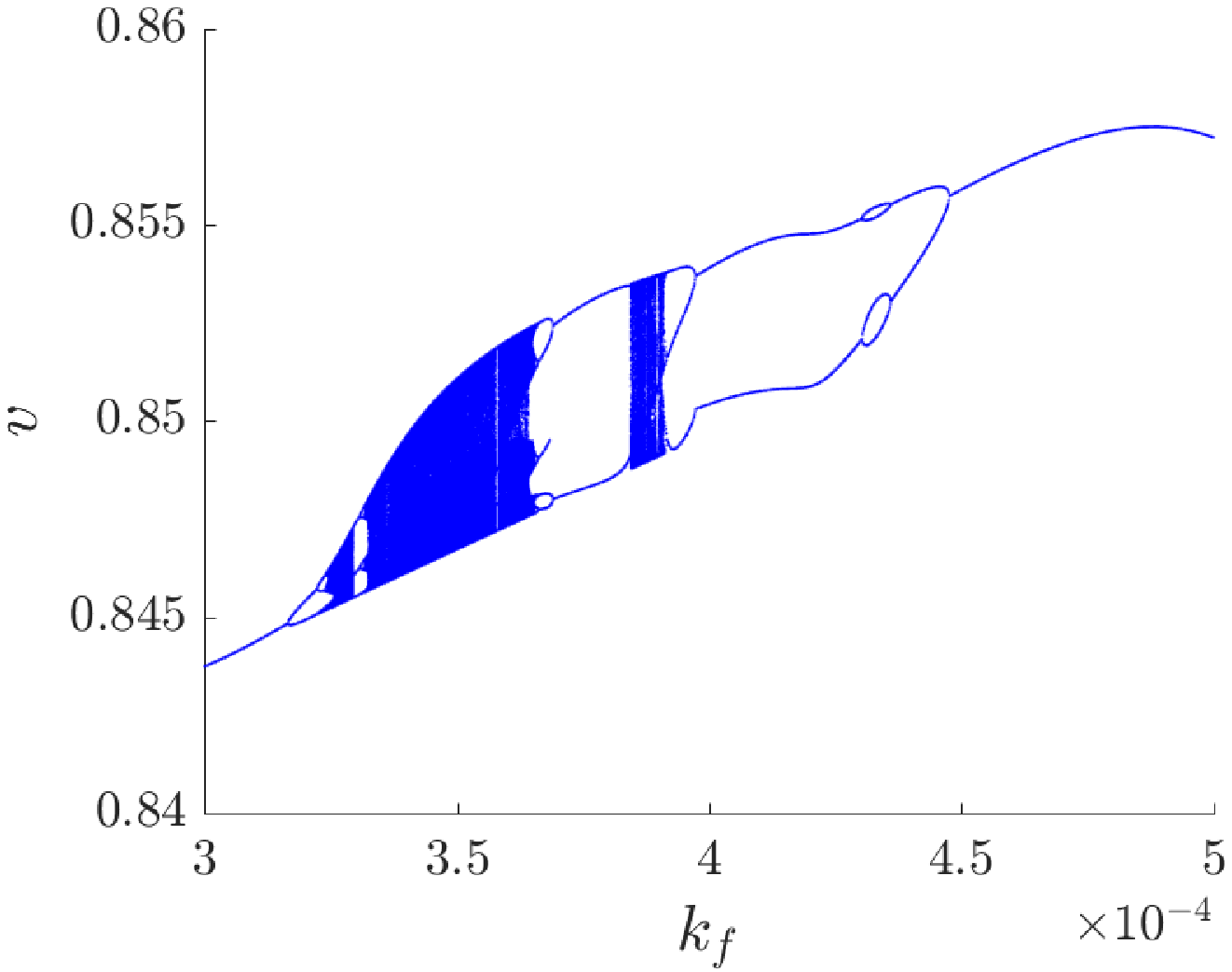}}
	\end{tabular}
	 \caption{Bifurcation diagram of the variable \(x\)(left), \(z\)(center), \(v\)(right) \(k_f \in [3 \times 10^{-4},5 \times 10^{-4}]\).}
	\label{bif}
\end{figure}

Next, the bifurcation diagram (constructed as a projection of a local maxima) of the model \eqref{all11} was plotted for each variable \(x\), \(z\) and \(v\) with respect to the free parameter \(k_f \in (3 \times 10^{-4}, 5 \times 10^{-4})\) in Figure \ref{bif}. 
In this bifurcation diagram, so-called "period doubling" and "windows" effects are also visible.
Periodic movement can be identified in the range of the parameter, 
e.g., \(k_f \in (3 \times 10^{-4}, 3.24 \times 10^{-4})\) and \(k_f \in (3.95 \times 10^{-4}, 5 \times 10^{-4})\). 
The interval in between these values is interrupted by chaotic cases around \(k_f = 3.25 \), 
and some \(k_f \in (3.34 \times 10^{-4}, 3.65 \times 10^{-4})\) and \(k_f \in (3.85 \times 10^{-4}, 3.9 \times 10^{-4})\).

\subsection{Approximate entropy}

The approximate entropy is a technique used to quantify the amount of regularity and unpredictability of fluctuations in time-series. 
The main advantages are that it can be computed on short time series and it allows to compare the differences in complexity of the same system with different parameters settings, see, e.g., \cite{Pincus1991} to be detected. 
More complex notion of entropy type can be found in, e.g., \cite{Threewayweightedcombinationentropiesbasedonthreelayergranularstructures}. 
To compute the approximate entropy, two parameters must be set: embedding dimension $m$ and neighborhood threshold $r$.
Let $s(t) \in \mathbb{R}$ for $t = \{1,2,\dots,N\}$ be a time series with $N$ observations.
Then embedded vector $S(t)$ at time $t$, is defined as $S(t) = [s(t), s(t+1),s(t+2),\dots,s(t+(m-1))]$, where $t$ is the observed time and $m$ is the embedding dimension.
The maximum distance of embedded vectors is computed as follows:
\begin{equation}
\label{eqn:maxdist}
D(i, j) = d(S(i),S(j)) = \max_{k=0,1,\dots,m-1}|s(i + k) - s(j + k)|,
\end{equation}
for $i,j = \{1,2,\dots,N-(m-1)\}$.

Compute the thresholded version of the distance with threshold given by $r$:
\begin{equation}
d_r(i,j) = 
\left\{\begin{array}{l@{\qquad}l}
1, & {}D(i,j) < r \\
0, & {}~otherwise, \\
\end{array}\right.
\end{equation}
for $i,j \in \{1,2,\dots, N-(m-1)\}.$

Compute $C_i^m(r)$ as a ratio between points in the neighborhood of i and the number of embedded vectors.
\begin{equation}
C_i^m(r) = \frac{\sum_{j=1}^{N-(m-1)} d_r(i,j)}{N-(m-1)}.
\end{equation}

Then compute the average of logarithm of all the $C_i^m(r)$
\begin{equation}
\Phi^m(r) = \frac{1}{N-(m-1)} \sum_{i=1}^{N-(m-1)}\ln C^m_i(r).
\end{equation}

Finally, approximate entropy for the finite time series with $N$ data points is computed as
\begin{equation}
ApEn(m,r,N) = \Phi^m(r) - \Phi^{m+1}(r).
\end{equation}
For robust estimation, it was suggested by Pincus \cite{Pincus1991} the time series is containing at least $10^3$ observations.

The approximate entropy was calculated using the \textit{TSEntropies} package \cite{apenbal} for R \cite{M2}. 
The computations were made for the input vector $s$ given in a normalized form of all state variables: 
$$s(t) = \sqrt{x^2(t)+z^2(t)+v^2(t)},$$
\(k_f\in(3 \times 10^{-4},5 \times 10^{-4})\) and \(r=0.1\). 
The results of approximate entropy for all values of the parameter \(k_f\) are in Figs.~\ref{np1}--\ref{np3}.

\subsection{0-1 test for chaos}

The 0-1 test for chaos, invented by Gottwald and Melbourne \cite{test}, is one of the methods for distinguishing between regular and chaotic dynamics of a deterministic system. In contrast to the other approaches, the nature of the system is irrelevant, thus the test can be applied directly onto experimental data, ordinary differential equations, or partial differential equations. The results return values close to either 0 or 1, with 0 corresponding to regular dynamics and 1 to chaotic dynamics. With its easy implementation, evaluation, and wide range of application, using this tool for detecting chaos is becoming more popular in different fields.
 
The 0-1 test for chaos can be computed by the following algorithm \cite{test}.

\begin{figure}
\centering
	\begin{tabular}{l r}
		\fbox{\includegraphics[scale=0.46]{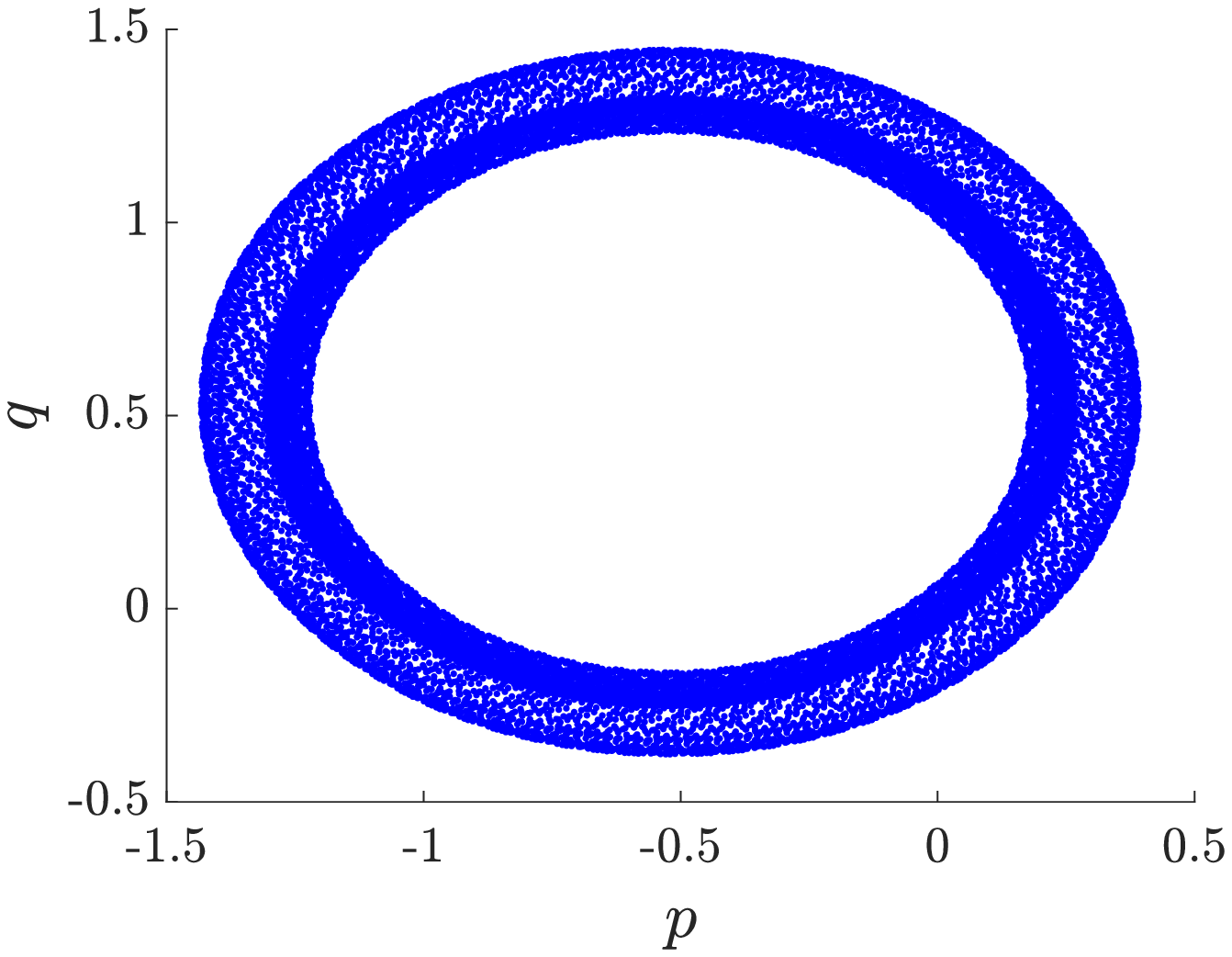}} & \fbox{\includegraphics[scale=0.46]{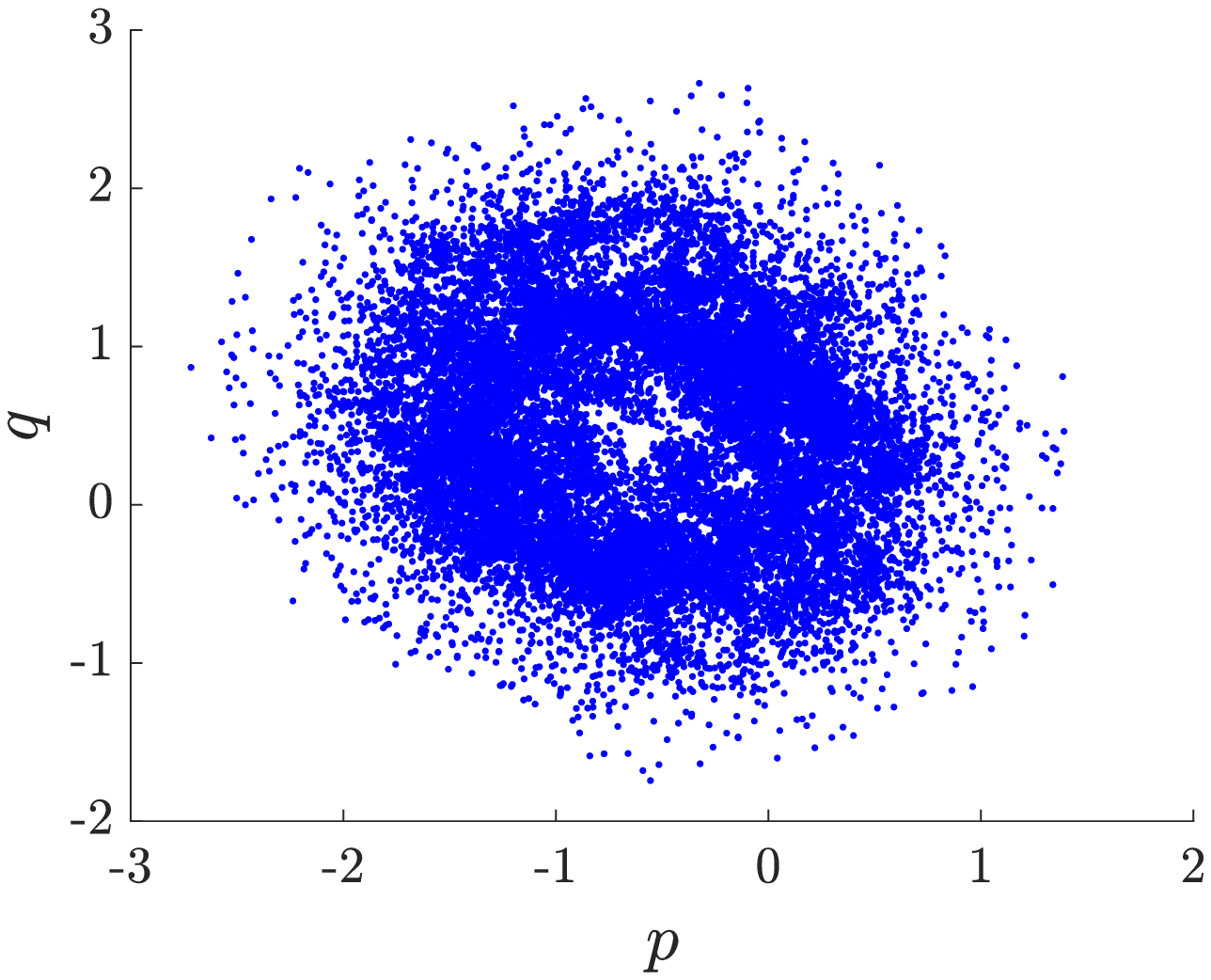}}
	\end{tabular}
	 \caption{The plot of \(p\) versus \(q\) for \(c=1.569853\) for \(k_f = 3\times 10^{-4}\) showing regular dynamics (left) and chaotic dynamics for \(k_f = 3.5\times 10^{-4}\)(right).}
	\label{pq}
\end{figure}

Given the observation \(\phi(j)\) for \(j=1,2,...,N\) and a suitable choice of \(c \in (0,2\pi)\), the following translation variables are computed:

\begin{equation*}
p_c(n)=\sum_{j=1}^{n}\phi(j)\cos (jc),
\end{equation*}
\begin{equation*}
q_c(n)=\sum_{j=1}^{n}\phi(j)\sin (jc)
\end{equation*}
 for \(n=1,2,...,N\). 
 The dynamics of the translation components $p_c$ and $q_c$ is shown on the  \(p_c\) versus \(q_c\) plot.
 A bounded trajectory is in Fig.~\ref{pq} (left) corresponding to regular movement, for \(k_f = 3\times 10^{-4}\).
 An unbounded trajectory is in Fig.~\ref{pq} (right) related to the chaotic case, for \(k_f = 3.5\times 10^{-4}\).

The idea for the 0-1 test, first described in \cite{test}, is that the boundedness or unboundedness of the trajectory \(\{(p_{j},q_{j})_{j\in[1,N]}\}\) can be studied through the asymptotic growth rate of its time-averaged mean square displacement (MSD), which is defined as
\begin{equation*}
M(n)=\lim_{N \to \infty}\frac{1}{N}\sum_{j=1}^{N}d(j,n)^{2}
\end{equation*}
where 
\begin{equation*}
d(j,n)=\sqrt{(p_{j+n}-p_{j})^{2}+(q_{j+n}-q_{j})^{2}}
\end{equation*}
is the time lapse of the duration \(n\) \((n\ll N)\) starting from the position at time \(j\). As it is shown in \cite{test,106}, it is important to use values of \(n\) small enough compared to \(N\), noted \(n_{cut}, (n \leq n_{cut})\). 
A subset of time lags \({n_{cut} \in [1,N/10]}\) is advised for the computation of each \(K_{c}\).

For bounded trajectories and regular dynamics, \( M(n)\) is a bounded function in time, whereas unbounded trajectories, meaning chaotic dynamics, are described by \( M(n)\) growing linearly with time. 
Thus the asymptotic growth rate of the MSD must be calculated, which correlates with the unboundedness of the trajectory.

As proposed in \cite{test}, the modified MSD is calculated as
\begin{equation*}
D(n)=M(n)-E(\phi)^{2}\frac{1-\cos (nc)}{1-\cos c}
\end{equation*}

The output of the 0-1 test for chaos is computed by the correlation method as
\begin{equation*}
K_{c}=\corr (\xi, \Delta) \in [-1, 1]
\end{equation*}
for the vectors \( \xi = (1, 2, ..., n_{cut}) \) and \( \Delta = (D_c(1), D_c(2), ..., D_c(n_{cut})) \).

The final result of the test is 
$$K= \median (K_{c}).$$

The position of the studied system (\ref{all11}) at any moment of time is determined by displacements $x$, $z$, and $v$, 
which are used for defining vector $\phi$:
$$\phi (j) = \sqrt{ x(j)^2 + z(j)^2 + v(j)^2 }.$$

\begin{figure}
\centering
	\begin{tabular}{l r}
		\fbox{\includegraphics[scale=0.46]{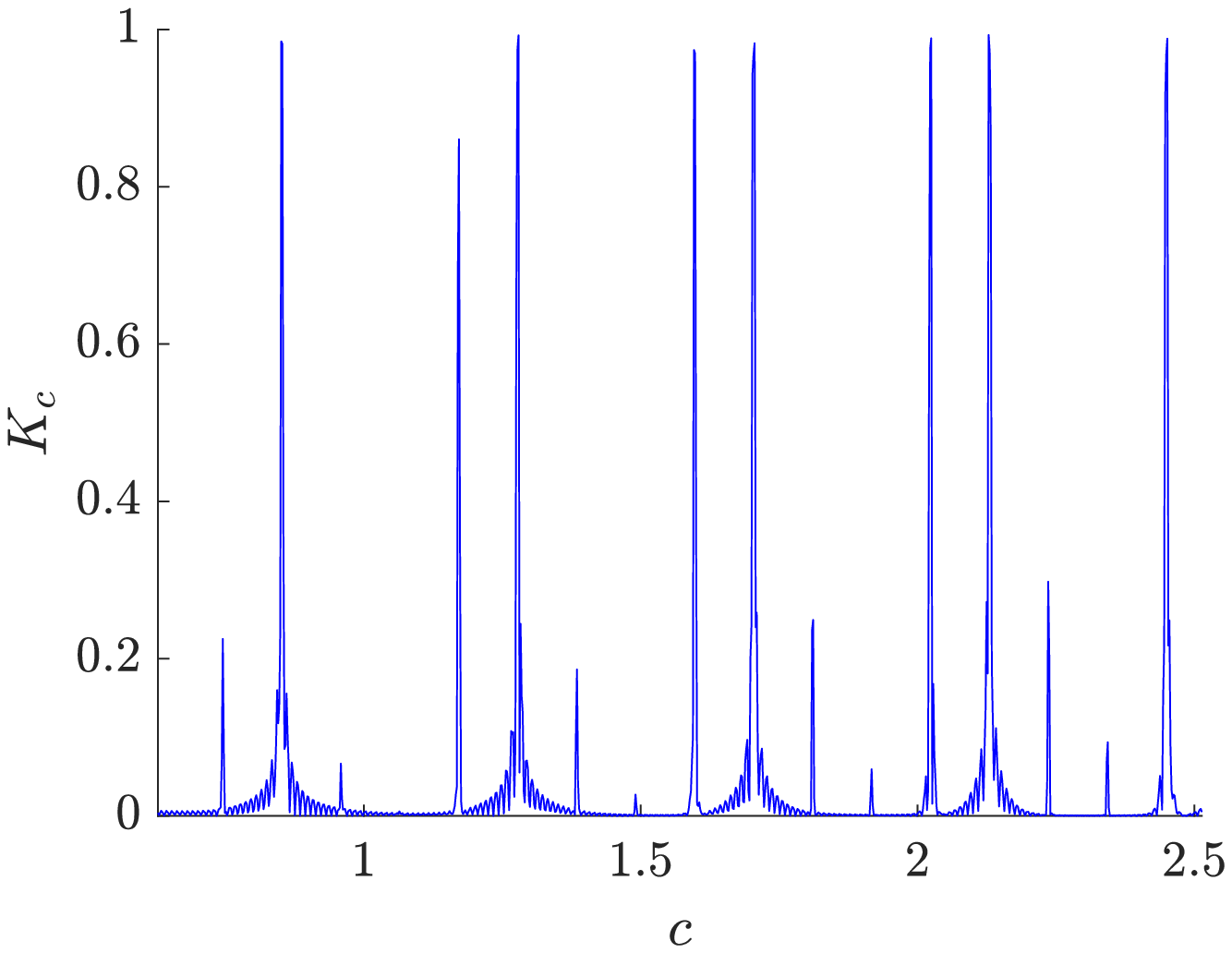}} & \fbox{\includegraphics[scale=0.46]{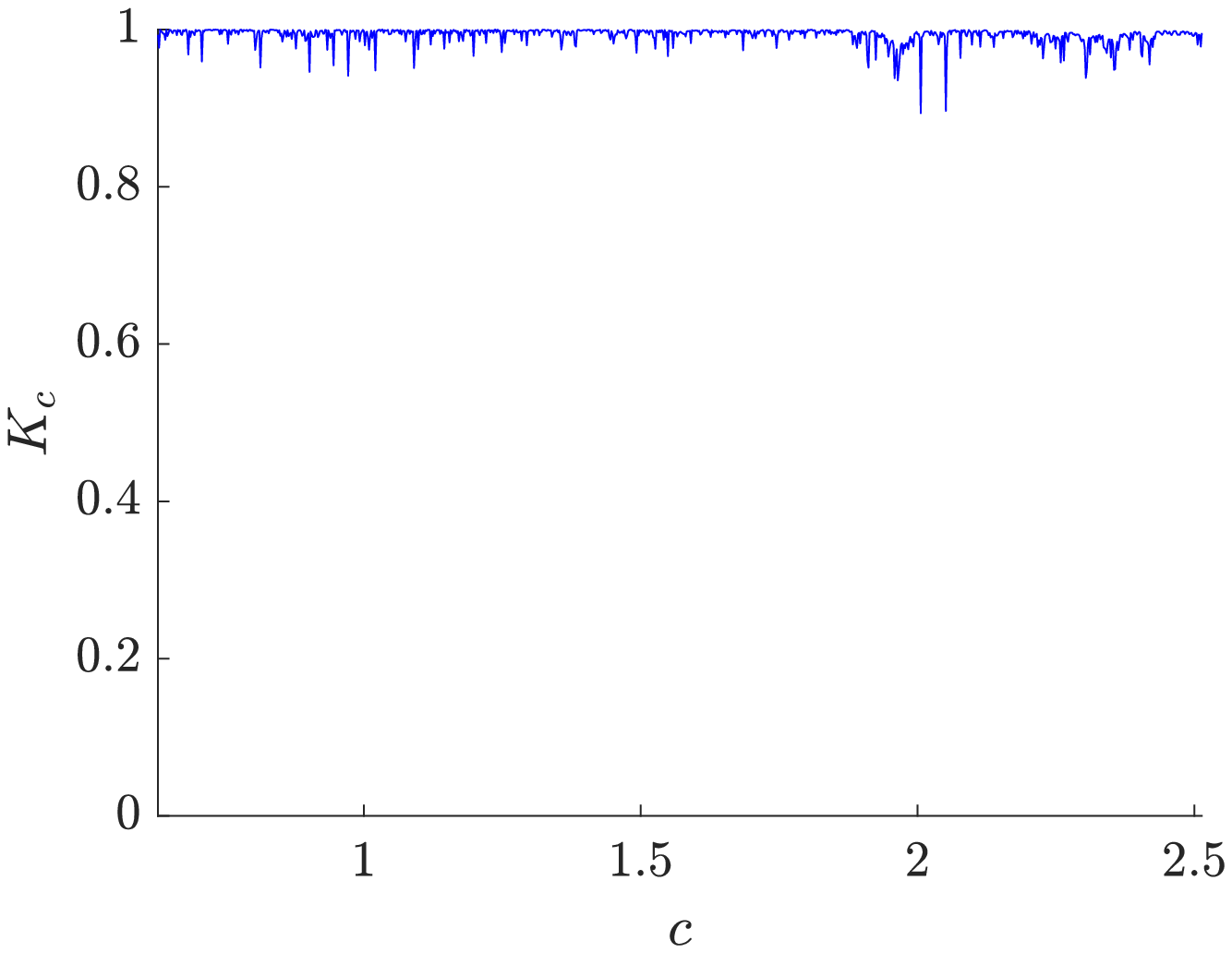}}
	\end{tabular}
	 \caption{The plot of \(K_c\) based on \(c\) for \(k_f = 3\times 10^{-4}\) showing regular dynamics (left) and for \(k_f = 3.5\times 10^{-4}\) showing chaotic dynamics (right).}
	\label{kc}
\end{figure}

For these simulations, a free software environment R \cite{M2} was used including the {\it Chaos01} package developed by T. Martinovi\v{c} \cite{M1}. 
Comparison of values \(K_c\) for periodic and chaotic case is shown in Fig.~\ref{kc}, for \(k_f = 3\times 10^{-4}\) and \(k_f = 3.5\times 10^{-4}\), respectively.

\begin{figure}
\centering
\fbox{\includegraphics[scale=0.45]{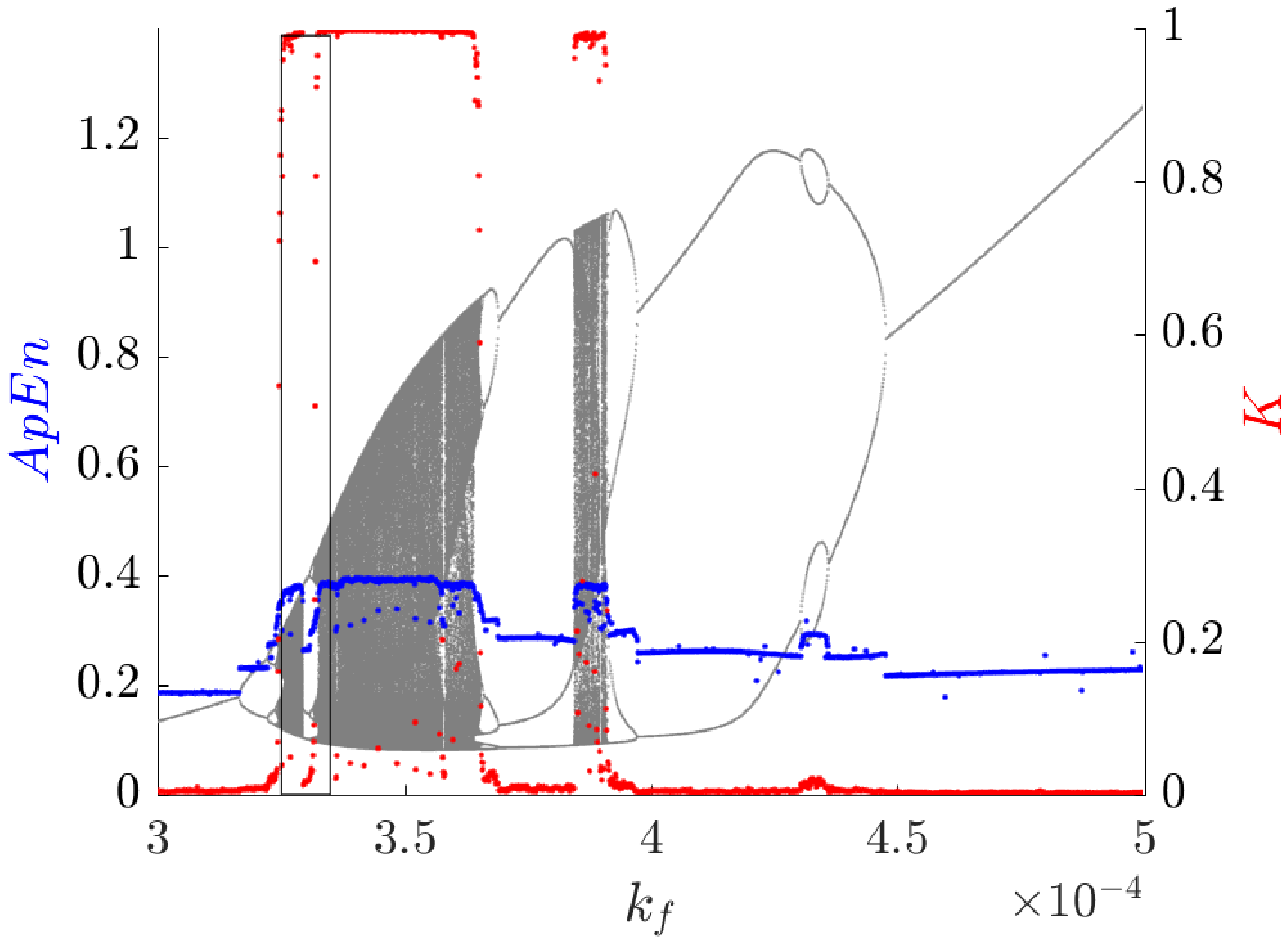}}
\caption{The result of approximate entropy (in blue) and the result of the 0-1 test for chaos (in red) for \(k_f \in (3\times 10^{-4},5\times 10^{-4})\). The magnification in the black rectangle is shown in Fig~\ref{np2}. The bifurcation diagram for variable \(x\) is shown in the background.}
\label{np1}
\end{figure}

\begin{figure}
\centering
\fbox{\includegraphics[scale=0.45]{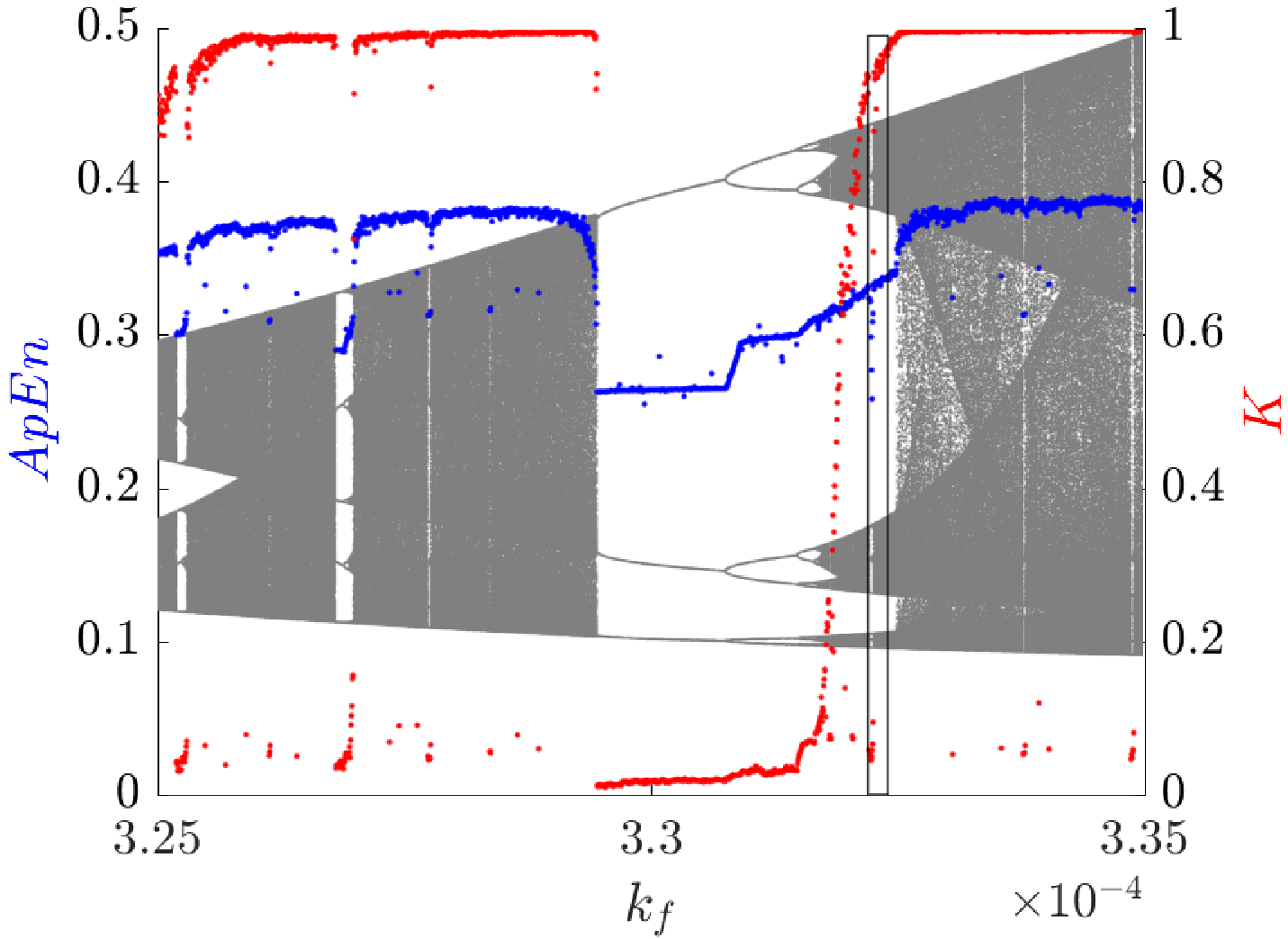}}
\caption{The result of approximate entropy (in blue) and the result of the 0-1 test for chaos (in red) for \(k_f \in (3.25\times 10^{-4},3.35\times 10^{-4})\). The magnification in the black rectangle is shown in Fig~\ref{np3}. The bifurcation diagram for variable \(x\) is shown in the background.}
\label{np2}
\end{figure}

\begin{figure}
\centering
\fbox{\includegraphics[scale=0.45]{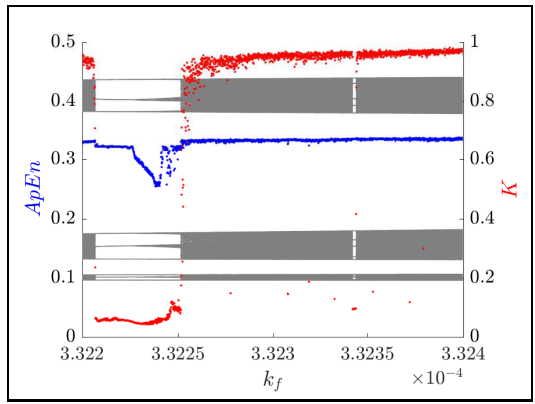}}
\caption{The result of approximate entropy (in blue) and the result of the 0-1 test for chaos (in red) for \(k_f \in (3.322\times 10^{-4},3.324\times 10^{-4})\). The bifurcation diagram for variable \(x\) is shown in the background.}
\label{np3}
\end{figure}

The results of the 0-1 test for chaos for all values of the parameter \(k_f\) are in Figs.~\ref{np1}--\ref{np3}. 

\section{Conclusions}\label{sec:concl}
In this paper, the GF model \eqref{all11} associated with the BZ chemical reaction was solved using adaptive six-stage, fifth-order, Runge-Kutta method implemented as \(ode45\) solver in Matlab. 
To eliminate the stiffness problem the model \eqref{all11} was also simulated by \(ode23s\) solver in Matlab \cite{Matlab}, outputs were identical.

The simulations were used to plot 3D phase portraits, bifurcation diagrams, the approximate entropy, and the 0-1 test for chaos. 
The mining process of dynamical properties were performed in the free software R \cite{M2} using packages \(TSEntopies\) \cite{apenbal} and \(Chaos01\) \cite{M1} depending on flow rate parameter $k_f$.

It is evident from  main results in Figs.~\ref{np1}--\ref{np3} that both tests clearly detect regular and irregular patterns for given $k_f$.
 
Our results show that the method of approximate entropy returns qualification constant which describes complexity in the system invariantly with respect to the origin.
On the other hand, the 0-1 test as qualification tool returns zero for regular (periodic or quasi-periodic) movement and one for irregular (chaos) characteristics. 

Moreover, if the output of the 0-1 test is not close to zero or one, then the examined test case has not yet reached attractor or reached intermittent state, see e.g. \cite{intermit1} or \cite{intermit2} and references therein.
 
Further, we observe a correlation between approximate entropy and the 0-1 test for chaos. In general, the increasing values of the 0-1 test for chaos are coupled to increasing approximate entropy and vice versa. 

We notice isolated low values of the 0-1 test for chaos accompanied by comparatively low values of approximate entropy well within of chaotic region characterized by high 0-1 test chaos values and approximate entropy. 
To investigate and zoom in, we have constructed three-stage system of nested subintervals of flow rates $k_f$, see Figs.~\ref{np1}--\ref{np3}, for which in every level the 0-1 test for chaos and approximate entropy was computed. 
At every level we observe the same pattern.
 
That naturally yields suggestion of a fractal structure in the set of $k_f$:

\begin{OpenProb}
Is there a totally disconnected (Cantor) set of flow rates $k_f$ in \([3\times 10^{-4}, 5\times 10^{-4}]\) such that for each such parameter the GF model \eqref{all11} is showing chaos?
\end{OpenProb}

\begin{OpenProb}
Is there a totally disconnected (Cantor) set of flow rates $k_f$ in \([3\times 10^{-4}, 5\times 10^{-4}]\) such that for each such parameter the GF model \eqref{all11} is showing regular movement?
\end{OpenProb}

\section*{Acknowledgement}
This work was supported by The Ministry of Education, Youth and Sports from the National Programme of Sustainability (NPU II) project ``IT4Innovations excellence in science -- LQ1602``;
 by The Ministry of Education, Youth and Sports from the Large Infrastructures for Research, Experimental Development and Innovations project ``IT4Innovations National Supercomputing Center -- LM2015070``; 
 by SGC grant No. SP2019/125 "Qualification and quantification tools application to dynamical systems", V\v{S}B - Technical University of Ostrava, Czech Republic,
Grant of SGS No. SP2019/84, V\v{S}B - Technical University of Ostrava, Czech Republic.

\end{document}